\documentclass[runningheads]{llncs}
\usepackage[T1]{fontenc}
\usepackage{color}
\usepackage{amsmath}
\usepackage{multirow}
\usepackage{multicol}
\usepackage{mathtools}
\usepackage{mathrsfs}
\usepackage{caption}

\usepackage{amsmath,amssymb,amsfonts}
\usepackage{algorithmic}
\usepackage{graphicx}
\usepackage{textcomp}
\usepackage{algorithm}
\usepackage{graphics}
\usepackage{threeparttable}
\usepackage{color}
\usepackage[normalem]{ulem}
\usepackage{multirow}
\usepackage{float}
\usepackage{amsfonts}
\usepackage{bm}
\usepackage{array}
\usepackage{colortbl}
\usepackage{pifont}
\usepackage{diagbox}
\usepackage{rotating}
\usepackage{booktabs}
\usepackage{overpic}
\usepackage{makecell}
\usepackage{contour}
\usepackage{adjustbox}
\usepackage[misc]{ifsym} 
\usepackage[T1]{fontenc}
\usepackage{graphicx}
\usepackage{xcolor}  
\usepackage{tabularray}
\usepackage{colortbl}
\usepackage{subcaption}
\usepackage{orcidlink}
\usepackage{todonotes}
\usepackage{amsmath}
\usepackage{hyperref}

\newcommand\blfootnote[1]{%
  \begingroup
  \renewcommand\thefootnote{}\footnotetext{#1}%
  \addtocounter{footnote}{-1}%
  \endgroup
}

\begin{document}
\title{Shortcut Learning in Medical Image Segmentation}

\titlerunning{Shortcut Learning in Medical Image Segmentation}
\authorrunning{Lin and Weng et al.}
\author{Manxi 
Lin\inst{1\dag}\textsuperscript{\orcidlink{0000-0003-3399-8682}},
Nina Weng\inst{1\dag}\textsuperscript{\orcidlink{0009-0006-4635-0438}}, 
Kamil Mikolaj\inst{1}\textsuperscript{\orcidlink{0000-0002-1631-9329}},
Zahra Bashir\inst{2,3}\textsuperscript{\orcidlink{0000-0002-2497-282X}},
\\ Morten B. S. Svendsen\inst{1,3}\textsuperscript{\orcidlink{0000-0002-4492-3750}},
Martin G. Tolsgaard\inst{3,4,5}\textsuperscript{\orcidlink{0000-0001-9197-5564}},
\\
Anders N. Christensen\inst{1}\textsuperscript{\orcidlink{0000-0002-3668-3128}},
Aasa Feragen\inst{1}\textsuperscript{(\Letter)}\textsuperscript{\orcidlink{0000-0002-9945-981X}}
}

\institute{
$^1$~Technical University of Denmark, Kongens Lyngby, Denmark\\
\email{\{manli, ninwe, afhar\}@dtu.dk}\\
$^2$~Slagelse Hospital, Slagelse, Denmark\\
$^3$~CAMES, Copenhagen, Denmark\\
$^4$ Copenhagen University Hospital Rigshospitalet, Copenhagen, Denmark\\
$^5$ Department of Clinical Medicine, University of Copenhagen, Copenhagen, Denmark
}

\maketitle             
\begin{abstract}
Shortcut learning is a phenomenon where machine learning models prioritize learning simple, potentially misleading cues from data that do not generalize well beyond the training set. While existing research primarily investigates this in the realm of image classification, this study extends the exploration of shortcut learning into medical image segmentation. We demonstrate that clinical annotations such as calipers, and the combination of zero-padded convolutions and center-cropped training sets in the dataset can inadvertently serve as shortcuts, impacting segmentation accuracy. We identify and evaluate the shortcut learning on two different but common medical image segmentation tasks. In addition, we suggest strategies to mitigate the influence of shortcut learning and improve the generalizability of the segmentation models. By uncovering the presence and implications of shortcuts in medical image segmentation, we provide insights and methodologies for evaluating and overcoming this pervasive challenge and call for attention in the community for shortcuts in segmentation. Our code is public at \href{https://github.com/nina-weng/shortcut\_skinseg}{https://github.com/nina-weng/shortcut\_skinseg}.  
\end{abstract}

\section{Introduction}
\blfootnote{
${}^\dag$ M. Lin and N. Weng contributed equally to this work.
}

\begin{figure*}[t]
\centering
  \includegraphics[width=.6\linewidth]{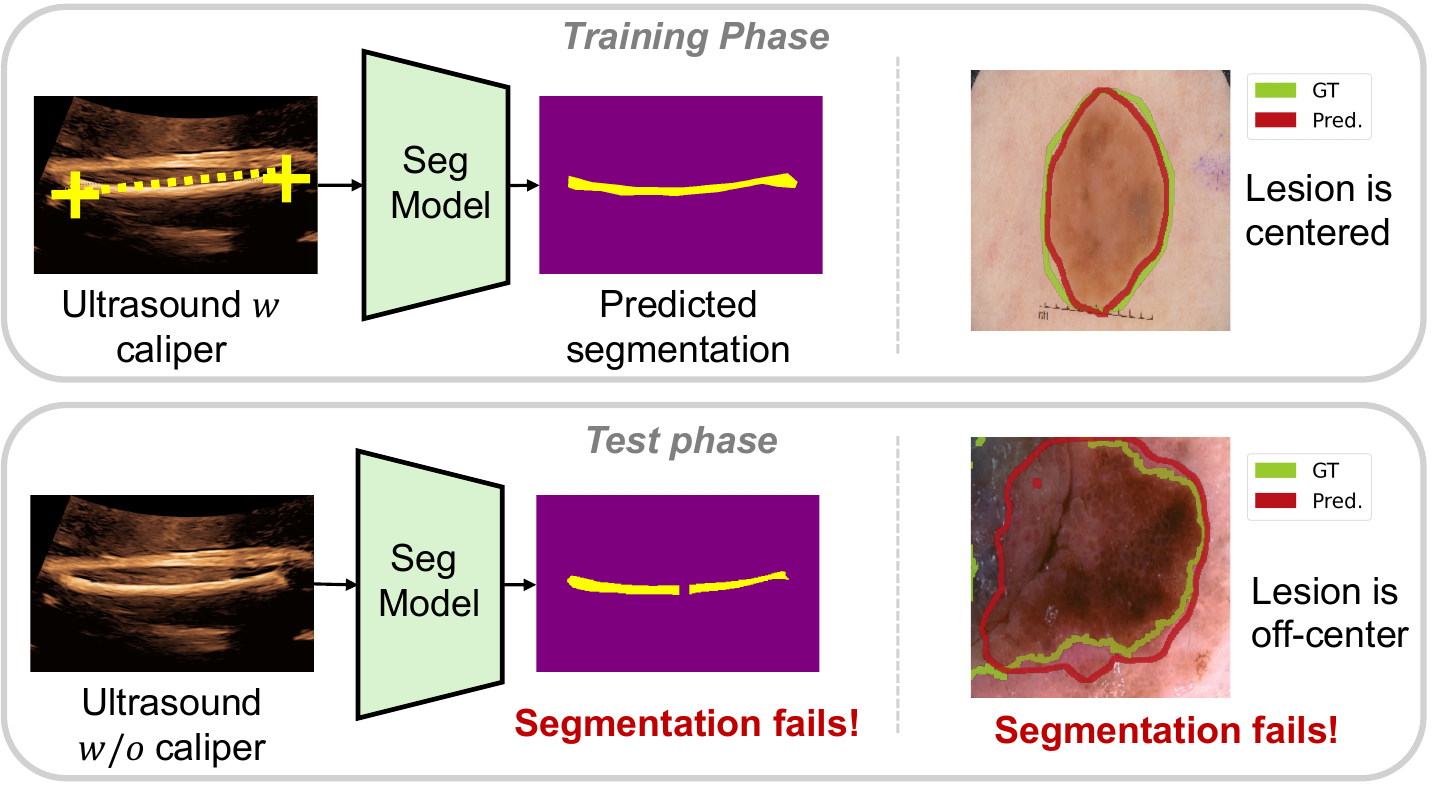}
\caption{Left: Shortcuts as yellow markers that affect ultrasound segmentation. The calipers were dilated for illustration. Right: Shortcuts that stem from the dataset construction, where lesions are center-cropped. }
    \label{fig:fetal}
\end{figure*}

Shortcut learning, a critical challenge in machine learning, refers to situations where models latch onto easy-to-learn correlations between input data and target features that are available at training time, but not necessarily at test time. This can result in models that perform well during development, but falter when exposed to real-world practice, where the ``shortcuts'' are absent. Shortcut learning receives increasing attention in medical imaging~\cite{narla2018automated,nauta2021uncovering}, where it affects high-stakes, high-risk clinical decisions~\cite{jimenez2023detecting}, undermines the trust in machine learning models~\cite{geirhos2020shortcut}, and even threatens model fairness~\cite{brown2023detecting}.

Most current work addresses shortcut learning in classification. Fig.~\ref{fig:fetal} shows an example of how clinical ultrasound screening images are, as a natural part of the clinical process, marked with text and measurement calipers at acquisition time. The text and calipers correlate strongly with specific anatomical standard planes, and as demonstrated in~\cite{mikolaj2023removing}, this correlation enables the model to use the markers rather than actual anatomical characteristics when classifying images into different anatomical planes. This can cause failure at test time, when e.g.~sorting images during an ultrasound examination before calipers are placed.

Shortcut learning has rarely been investigated in other tasks such as image segmentation -- an essential medical image quantification task that demands exceptional accuracy and generalizability. Intuitively, shortcut learning might seem less of a risk for image segmentation: In contrast to classification, segmentation necessitates a precise delineation of organ boundaries. This task utilizes spatial relationships and textures within the image, as opposed to standalone, identifiable markers. As a result, it might seem unrealistic that a segmentation model affected by shortcuts could segment anything at all~\footnote[1]{See e.g.\url{https://gwern.net/tank}, point 6 under “Could it happen”)}. However, this only addresses the danger of false positives.

\textbf{In this paper, we challenge this prevailing assumption and show that shortcut learning can also affect medical image segmentation models.} We show how shortcut learning can lead to false negatives (failure to segment) at test time, greatly affecting segmentation performance.

We showcase two types of shortcuts in segmentation, with different underlying mechanisms, in fetal ultrasound and dermatological segmentation tasks. The first type mirrors the shortcuts identified in classification tasks. The other arises when center-cropped images are zero-padded in CNNs. As a result, the models learn that the objects to be segmented are located in the middle of the image, and they therefore fail to recognize objects near the border. This is important since center cropped patches are a very common segmentation dataset design, also in MICCAI challenges~\cite{wang2023pymic,bao2023boston}, see also Fig.~\ref{fig:fetal}.

\section{Background and Related Work}
A wide range of variables have been shown to act as shortcuts in medical images. In a large collection of examples, the model \textbf{learns signs of treatment, diagnosis notes, or signs of related diseases} as proxies of the target $y$. Known examples include chest tube in chest x-ray as a common treatment for Pneumothorax \cite{jimenez2023detecting}; clinical annotations or diagnosis notes which encode clinicians' suspected diagnoses, such as ink circling suspicious skin lesions or rulers measuring their size~\cite{narla2018automated,nauta2021uncovering}; text and calipers in ultrasound \cite{mikolaj2023removing} annotated primarily on high-quality images; or effects of disease symptoms such as movement artifact in MRI scans, which can appear as a result of tremor associated with Parkinson’s disease, or decreased capability of following instructions of staying still due to diseases such as dementia~\cite{godenschweger2016motion}.

In a different class of shortcuts, the \textbf{model encodes global information about the data collection process},
such as a hospital or hardware information, e.g different hospitals might use different scanners and protocols, which give the images a distinctive 'style'~\cite{glocker2019machine,zech2018variable}; data collected from different hospitals might have different prevalence distributions, e.g. a cancer-specialized hospital might have more samples of diseased subjects than a community hospital. 

\textbf{Model mistakenly learns the demographic features of the patient, which is associated with the disease prevalence}. Studies have shown that models can easily learn the race or gender information from chest X-rays \cite{glocker2023algorithmic}, increasing the concerns of using demographic features for prediction, especially when prevalence varies among groups.

\textbf{Shortcuts performance differs by the choice of model structures.} Studies show that model structures and loss functions could impact shortcut learning, e.g., 
Izmailov and Kirichenko et al. ~\cite{izmailov2022feature} studied how different choices of model architecture and pre-training strategy impact shortcut learning.
   
Whereas most case studies are classification tasks, in this paper, we showcase that shortcuts can also affect medical image segmentation.

\section{Shortcut A: Calipers and texts in fetal ultrasound} \label{sec:case_a}
Organ segmentation in fetal ultrasound is crucial for measurements and guiding predictive models. Segmentation models are typically trained on clinical screening images, which often include clinical annotations like text and measurement calipers specific to each anatomical plane~\cite{khalil2024isuog}. These annotations are absent during real-time applications, such as highlighting organs during scans. Here, we document how text and calipers serve as shortcuts for ultrasound segmentation.

\subsection{Experiments and observations}
\subsubsection{Data and experiment settings}
Our study involved segmenting organs from 3,775 third-trimester ultrasound images derived from a proprietary fetal ultrasound screening dataset. These images represent 13 distinct fetal anatomical structures and originate from four specific anatomical views: the fetal head, femur, abdomen, and the maternal cervix. A medical doctor annotated each image, with an experienced expert conducting subsequent verification. We partitioned the data into training, validation, and test sets, maintaining an 80/10/10 percent split. 
We utilized the official implementation of DTU-Net~\cite{lin2023dtu}. We trained the model with the AdamW optimizer for 1,000 epochs with weight decay 1e-6, batch size 32, and early stopping. Training began with a learning rate of 1e-4 and was adjusted by a cosine annealing schedule.

\begin{figure}[t]
    \centering
    \includegraphics[width=0.8\linewidth]{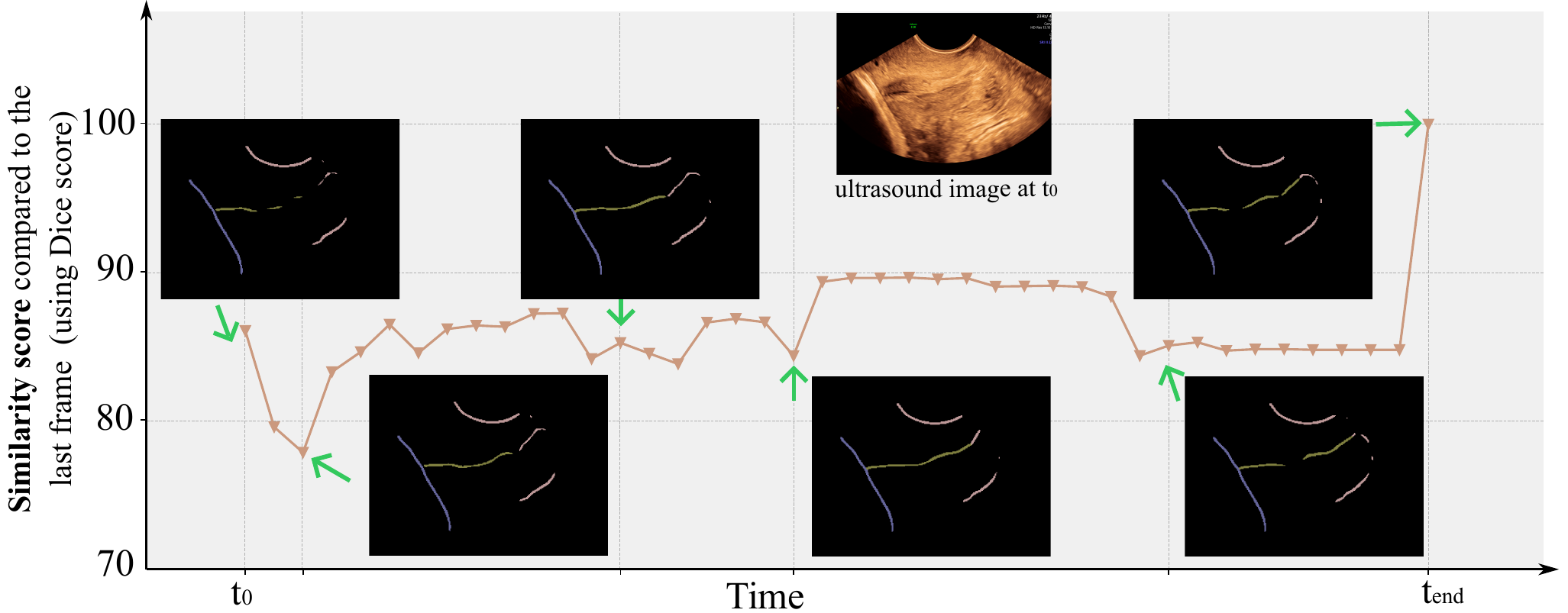}
        \caption{Illustration of the dynamic variation of model segmentation prediction on the maternal cervix in a video. The similarity score between frames at time $t_0$ and the last frame $t_{end}$ is computed by dice score. The selected frame shows that the segmentation prediction is unstable when putting the calipers and texts.
        }
    \label{fig:fetal_curve}
\end{figure}
\begin{table}[b]
    \centering
    \captionof{table}{Model performance on a test set with clinical annotations, and with the same annotations artificially removed. We report the average and standard deviation of the per-image Dice coefficient.
    }
    \begin{tabular}{l|c|c|c|c}
    \hline
    \multirow{2}{*}{Image} & \multicolumn{4}{c}{Anatomical plane}\\
    \cline{2-5}
     ~  & Head & Abdomen & Femur & Cervix \\
   \hline
       Test set \textit{w} annotation & $76.97\pm5.10$ & $82.06\pm6.60$ & $93.82\pm1.79$ & $76.29\pm4.05$\\
       \hline
      Test set \textit{w/o} annotation & $70.85\pm8.24$ & $78.85\pm7.72$ & $91.84\pm4.87$ & $71.81\pm4.86$\\
    \hline
    \end{tabular}
    \label{tab:fetal_performance}
\end{table}

\subsubsection{Demonstrating shortcut learning}
We first validate model performance on two versions of the test set. First, we use original images with text and calipers, and second, a ``clean'' version of the same test set where annotations are artificially removed via the inpainting strategy from~\cite{mikolaj2023removing}. Tab.~\ref{tab:fetal_performance} shows a performance drop on the clean images, indicating that annotations are used as shortcuts.

We further evaluate the impact of these shortcuts in real-time scans in the clinics. Here, we analyze 110 screen capture video recordings of ultrasound examinations from a single hospital, where the sonographer sequentially identified the anatomical structures of the head, abdomen, and femur in 100 videos, and performed cervix examination in the remaining 10 videos. Upon locating each anatomical feature, the image was frozen and then annotated with texts and calipers. We observe fluctuations in the segmentation predictions throughout this process, even when the underlying ultrasound image is frozen, see Fig.~\ref{fig:fetal_curve}, and a video clip example can be found in the supplementary material. The first row of Tab.~\ref{tab:fetal_variation} shows the impact of shortcuts on all 110 videos. We quantify segmentation robustness by reporting the difference in average dice coefficients of the segmentation predictions on the frames at the start and the end of the annotation process, i.e., the same image with and without clinical notes.

\begin{table}[t]
    \centering
    \captionof{table}{Model performance varies during the recorded annotation of a frozen ultrasound plane. The variation is measured with average dice coefficients (\%). `Baseline' is the model that suffers from shortcuts while `Mitigation' is the model trained on `clean' images. The mitigated model is better and far more stable.
    }
    \begin{tabular}{l|c|c|c|c}
    \hline
    \multirow{2}{*}{Model} & \multicolumn{4}{c}{Anatomical plane}\\
    \cline{2-5}
     ~  & Head & Abdomen & Femur & Cervix \\
   \hline
       Baseline & $89.07\pm14.9$ & $94.75\pm12.6$ & $91.51\pm19.6$ & $82.42\pm10.1$\\
       \hline
 \rowcolor[RGB]{240,240,240}     Mitigation & $98.39\pm1.3$ & $98.53\pm1.9$ & $95.38\pm8.9$ & $97.00\pm1.2$\\
    \hline
    \end{tabular}
    \label{tab:fetal_variation}
\end{table}

\subsubsection{Mechanism}
We are inspired by the conclusion from~\cite{mikolaj2023removing} for image recognition. In our segmentation task, the shortcut arises from the consistent co-occurrence of anatomical structures and the calipers and texts within our training dataset. This co-occurrence leads our segmentation model to rely on these clinical notes as predictive shortcuts, rather than fully understanding the texture and shape features of the anatomical structures themselves. 
This reliance undermines the model's performance in applications, posing a risk to its clinical applicability.

\subsubsection{Mitigation}
Following~\cite{mikolaj2023removing}, we trained a mitigated model on images with annotations artificially removed. As shown in the second row of Tab.~\ref{tab:fetal_variation}, performance increases substantially, indicating that text and calipers acted as shortcuts.

\section{Shortcut B: The combination of zero-padded convolutions and center cropped training sets}

Skin lesions refer to damaged or abnormal skin tissue that differs in appearance from the surrounding area. Common causes of skin lesions include abscesses, cysts, tumors, sores, and ulcers \cite{NCI_lesion}. Accurately segmenting skin lesions is critical for various downstream tasks, such as dermoscopic feature detection and lesion classification, particularly for detecting malignant skin cancers.

Here, we utilized the widely-used ISIC skin lesion segmentation dataset~\cite{codella2018skin} to illustrate our second case of segmentation shortcut learning, which stems from the combination of two typical design choices in image segmentation: \textbf{(i)} In skin lesion segmentation datasets, 
regions of interests, i.e., lesions tend to be shown quite close to the center of the image. This has been dubbed the 'common photographer bias' in computer vision~\cite{kirillov2023segment} but is also prevalent in popular medical datasets, see Fig.~\ref{fig:segmented_object_centre_dist_medical_dataset}.
\textbf{(ii)} A very common modeling choice is to use zero padding in CNN segmentation architectures to create output predictions with the same shape as the input image. In addition to ease of implementation, this also enables a full segmentation, rather than the partial segmentation of the image that a CNN without padding would deliver. 

\begin{figure}[t]
\setlength{\tabcolsep}{2pt}
\centering
\scalebox{0.75}{
\begin{tabular}{cccccccc}
  \centering

          \rotatebox{90}{\footnotesize Original}

 & \includegraphics[width=0.12\linewidth]{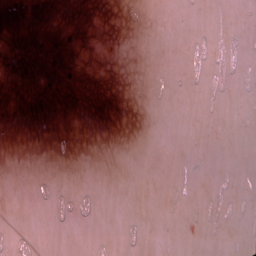}
 &\includegraphics[width=0.12\linewidth]{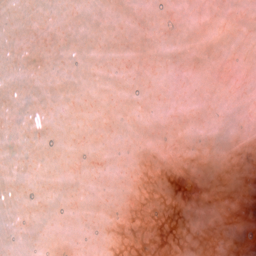}
 &\includegraphics[width=0.12\linewidth]{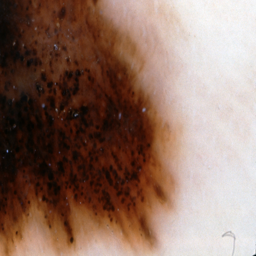}
&\includegraphics[width=0.12\linewidth]{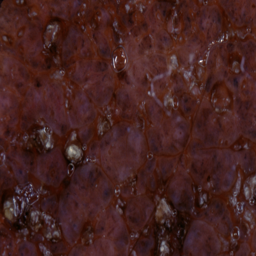}
&\includegraphics[width=0.12\linewidth]{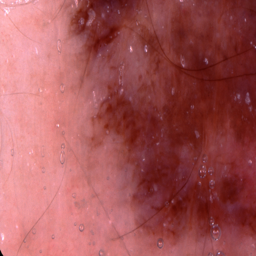}
&\includegraphics[width=0.12\linewidth]{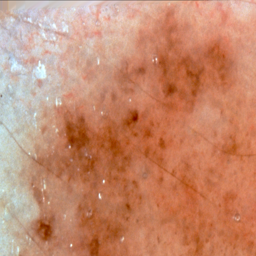}
&\includegraphics[width=0.12\linewidth]{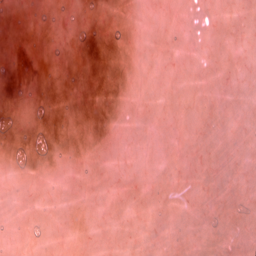}

\\

  \rotatebox{90}{\footnotesize GT}
&\includegraphics[width=0.12\linewidth]{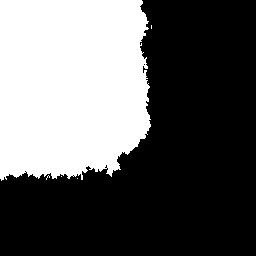}
&\includegraphics[width=0.12\linewidth]{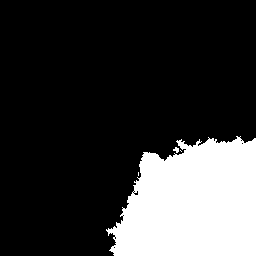}
   &\includegraphics[width=0.12\linewidth]{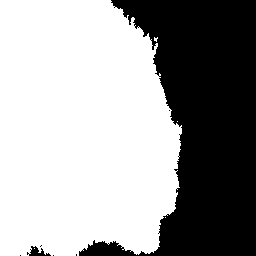}
   &\includegraphics[width=0.12\linewidth]{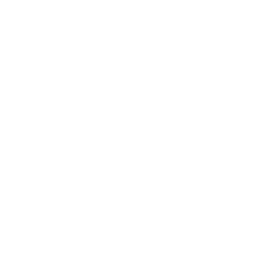}
   &\includegraphics[width=0.12\linewidth]{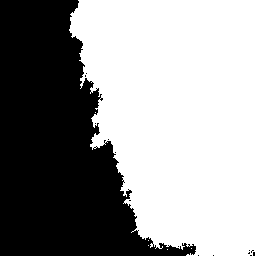}
   &\includegraphics[width=0.12\linewidth]{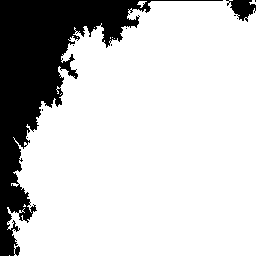}
   &\includegraphics[width=0.12\linewidth]{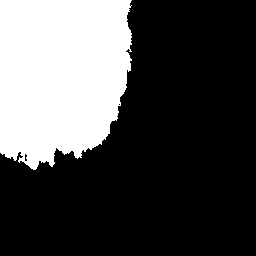}
   
   \\

 \rotatebox{90}{\footnotesize $M_{ori}$}

   &\includegraphics[width=0.12\linewidth]{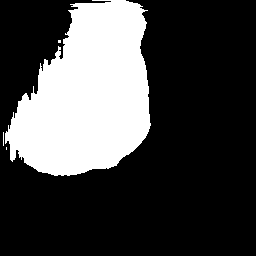} 
      &\includegraphics[width=0.12\linewidth]{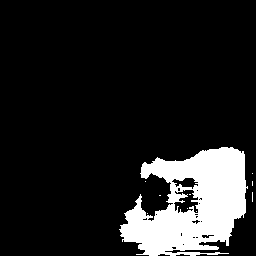} 
 &\includegraphics[width=0.12\linewidth]{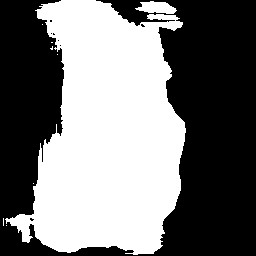}
  &\includegraphics[width=0.12\linewidth]{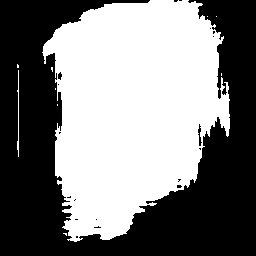} 
  &\includegraphics[width=0.12\linewidth]{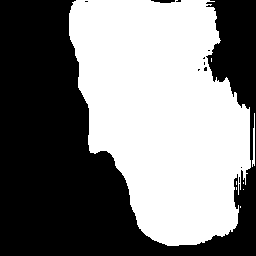}
  &\includegraphics[width=0.12\linewidth]{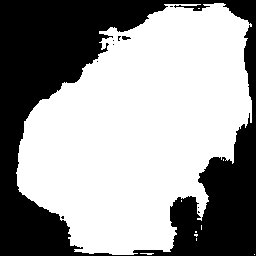}
  &\includegraphics[width=0.12\linewidth]{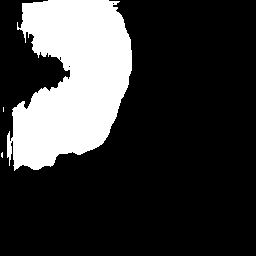}
  
  \\
  
      \rotatebox{90}{\footnotesize $M_{crop}$}
   &\includegraphics[width=0.12\linewidth]{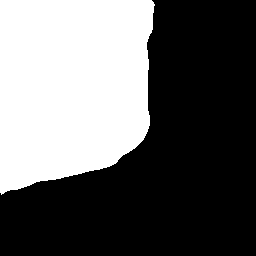}
   &\includegraphics[width=0.12\linewidth]{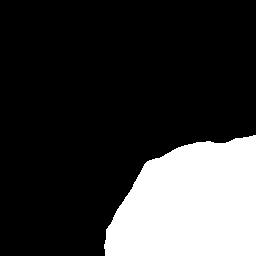}
   &\includegraphics[width=0.12\linewidth]{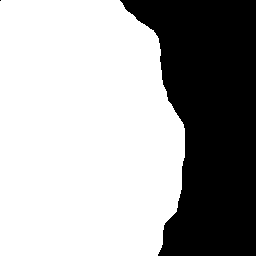} 
   &\includegraphics[width=0.12\linewidth]{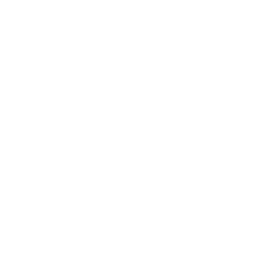} 
   &\includegraphics[width=0.12\linewidth]{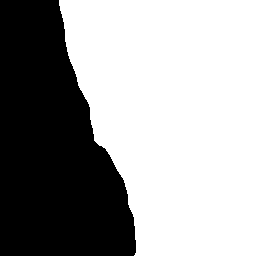}
   &\includegraphics[width=0.12\linewidth]{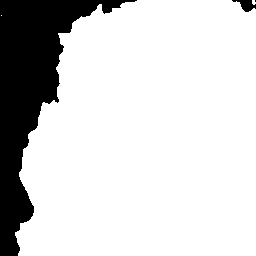}
   &\includegraphics[width=0.12\linewidth]{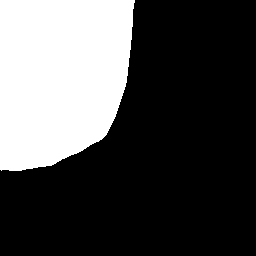}

\end{tabular}
}
   \caption{Sample lesion images from $\mathbf{ISIC_{CROP}}$ with ground truth (GT) segmentation and predicted segmentation by $M_{ori}$ and $M_{crop}$. We clearly see shortcut learning and its mitigation when training on $M_{ori}$ and $M_{crop}$, respectively.}
   \label{fig:samples_cropped_exp}
\end{figure}

However, the above two design choices combined lead to an unfortunate correlation between the features of pixels near the boundary and their class (typically background): For any pixel whose receptive field extends beyond the original image, its receptive field will contain a band of constant zero values (Fig.~\ref{fig:zeropad}), and pixels with such features will -- due to the centering of lesions in the dataset -- almost always belong to the background class. In other words, the combination of zero padding and dataset construction constitutes a shortcut that lets the model learn that pixels near the image boundary should be segmented as background. 

\begin{figure}[t]
    \centering
\scalebox{0.9}{
      \begin{subfigure}[t]{.26\textwidth}
  \centering
  \includegraphics[width=.99\linewidth]{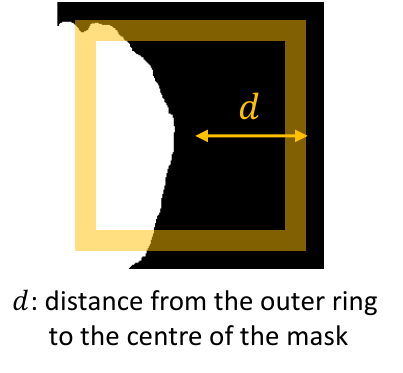}
  \caption{Ring area given $d$.}
  \label{fig:dice_ring_area_illu_d}
\end{subfigure}
~
    \begin{subfigure}[t]{.35\textwidth}
          \centering
          \includegraphics[width=.99\linewidth]{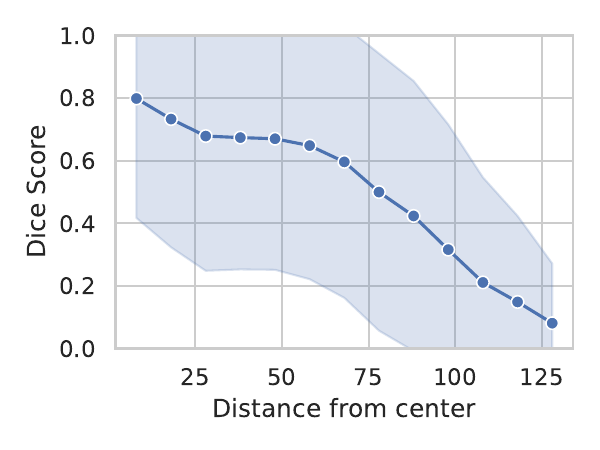}
          \caption{Performance of $M_{ori}$.}
          \label{fig:dice_ring_area_ori}
        \end{subfigure}%
~
    \begin{subfigure}[t]{.35\textwidth}
      \centering
      \includegraphics[width=.99\linewidth]{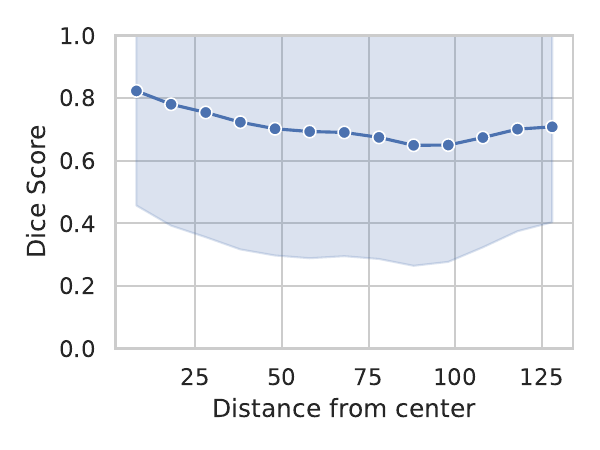}
      \caption{Performance of $M_{crop}$.}
      \label{fig:dice_ring_area_crop}
     \end{subfigure}
     }
    \caption{To emphasize the effect of shortcut learning, we separate the image into yellow bands of increasing distance from the center (a) and compute segmentation Dice scores within each region. Note how the Dice score of the shortcut model $M_{ori}$ declines with increasing distance from the center (b), while the mitigated model $M_{crop}$ maintains its segmentation ability across the image (c).}
    \label{fig:dice_ring_area}
\end{figure}

\subsection{Experiments and observations}

\subsubsection{Data and experimental settings} We use the \textbf{ISIC2017} skin lesion dataset~\cite{codella2018skin} with its original train and test splits. To show the effect of shortcut learning via center cropping and zero padding, we create an asymmetrically cropped version, $\mathbf{ISIC_{CROP}}$, where the original images are cropped into 4 quarters, one of which is randomly selected (see Fig.~\ref{fig:samples_cropped_exp}). Segmentations were made with a standard UNet~\cite{ronneberger2015u} trained with Dice loss \cite{sudre2017generalised} using Adam~\cite{kingma2014adam} with a learning rate of 0.001, 100 epochs and batch size 16. All images and masks are resized to $256 \times 256$.

\subsubsection{Demonstrating shortcut learning} 
Segmentation models $M_{ori}$ and $M_{crop}$ were trained using the original train splits within \textbf{ISIC2017} and $\mathbf{ISIC_{CROP}}$, respectively. As illustrated in Fig~\ref{fig:samples_cropped_exp}, $M_{ori}$ has problems segmenting lesions near the boundary. This is further quantified in Fig.~\ref{fig:dice_ring_area} (a), where we show how segmentation performance decreases in regions near the image boundary.

\begin{figure}[t]
\includegraphics[width=\linewidth]{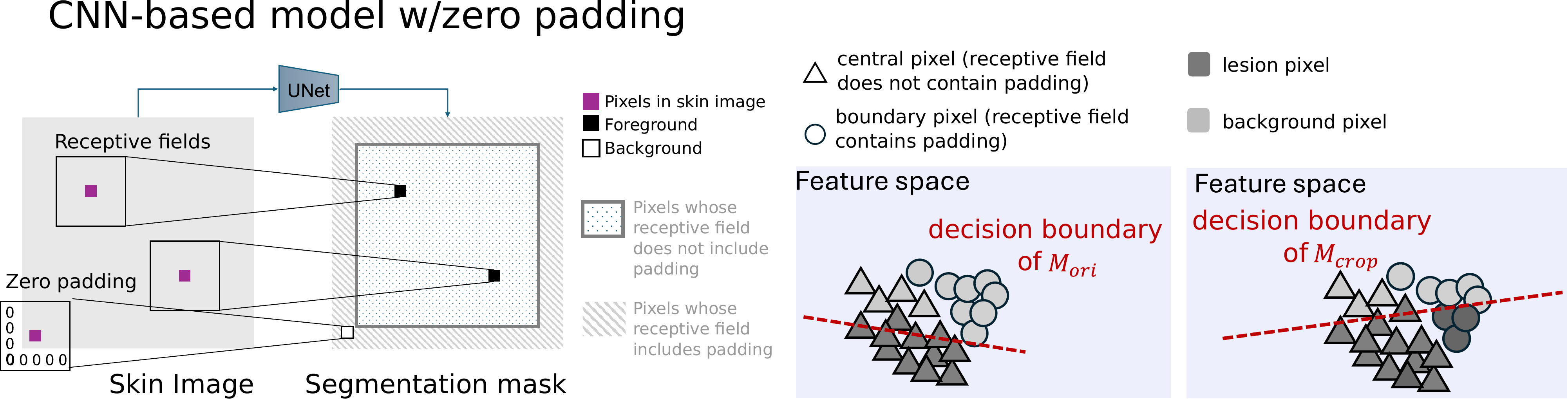}
\centering
\caption{Zero padding introduces a change in the features of segmented pixels whose receptive field contains padding -- i.e.~those that are near the boundary. This affects their encoding in feature space.}
\label{fig:zeropad}
\end{figure}
\subsubsection{Mechanism} To understand why zero padding causes shortcut learning, we need to return to the basics of CNNs: CNN segmentation networks are translation equivalent, meaning that without zero padding, the CNN does not encode any information about the spatial location of individual pixels. However, the feature patterns induced by padding enable the CNN to encode -- in a band around the image border -- the distance to the image border. As shown in Fig.~\ref{fig:zeropad}, this likely leads to these pixels clustering in feature space due to their special zero padding features. When using the \textbf{ISIC2017} dataset to train $M_{ori}$, those pixels almost exclusively belong to the background, which encourages the model to simply classify all boundary pixels as background.

\subsubsection{Mitigation} There is a simple yet efficient mitigation strategy: Namely to \emph{not} center crop training images (equivalently, augment with random cropping). This is documented by the $M_{crop}$ model, see Fig.~\ref{fig:samples_cropped_exp} and~\ref{fig:dice_ring_area} (c). When lesions are no longer centered during training, test performance no longer depends on the distance to the boundary, and the shortcut has been mitigated. Indeed, while $M_{ori}$ can barely categorize any edging pixels into the class of skin lesion, $M_{crop}$ clearly accepts the boundary of the image being the skin lesion.

An alternative mitigation could be to swap zero padding with reflection or replication. As these would still have distinctive features, shortcut learning would still be possible, but likely less pronounced as the shortcuts would be more difficult to learn. Nevertheless, crop augmentation is preferable as it completely removes the correlation between boundary affinity and segmentation labels.

\section{Discussion and conclusion}

\begin{figure}[t]
    \centering
\scalebox{0.8}{
\begin{tabular}{cccc}
  \centering
  ISIC2017~\cite{codella2018skin} & LIDC-IDRI~\cite{armato2011lung} & BraTS~\cite{menze2014multimodal} & Prostate~\cite{antonelli2022medical}
  \\
  \includegraphics[width=0.18\linewidth] {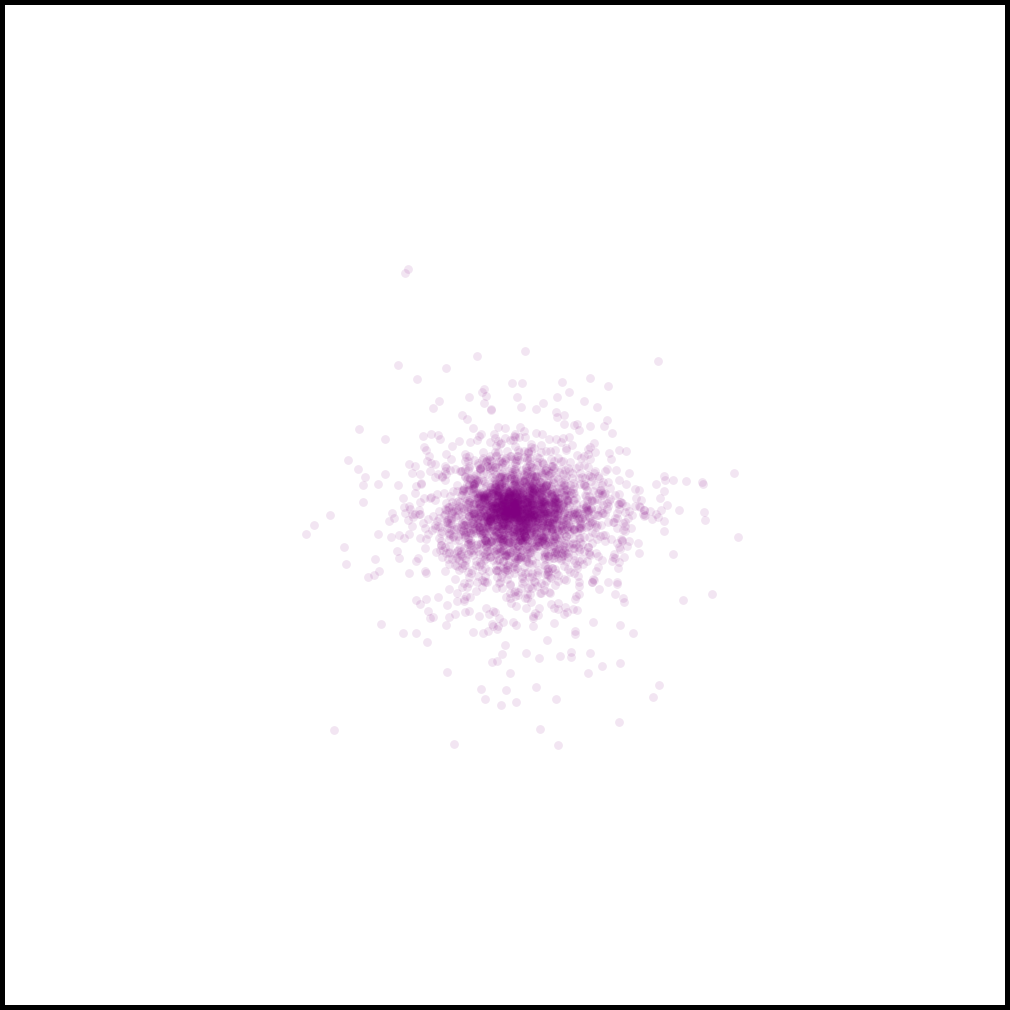} 
  &\includegraphics[width=0.18\linewidth]{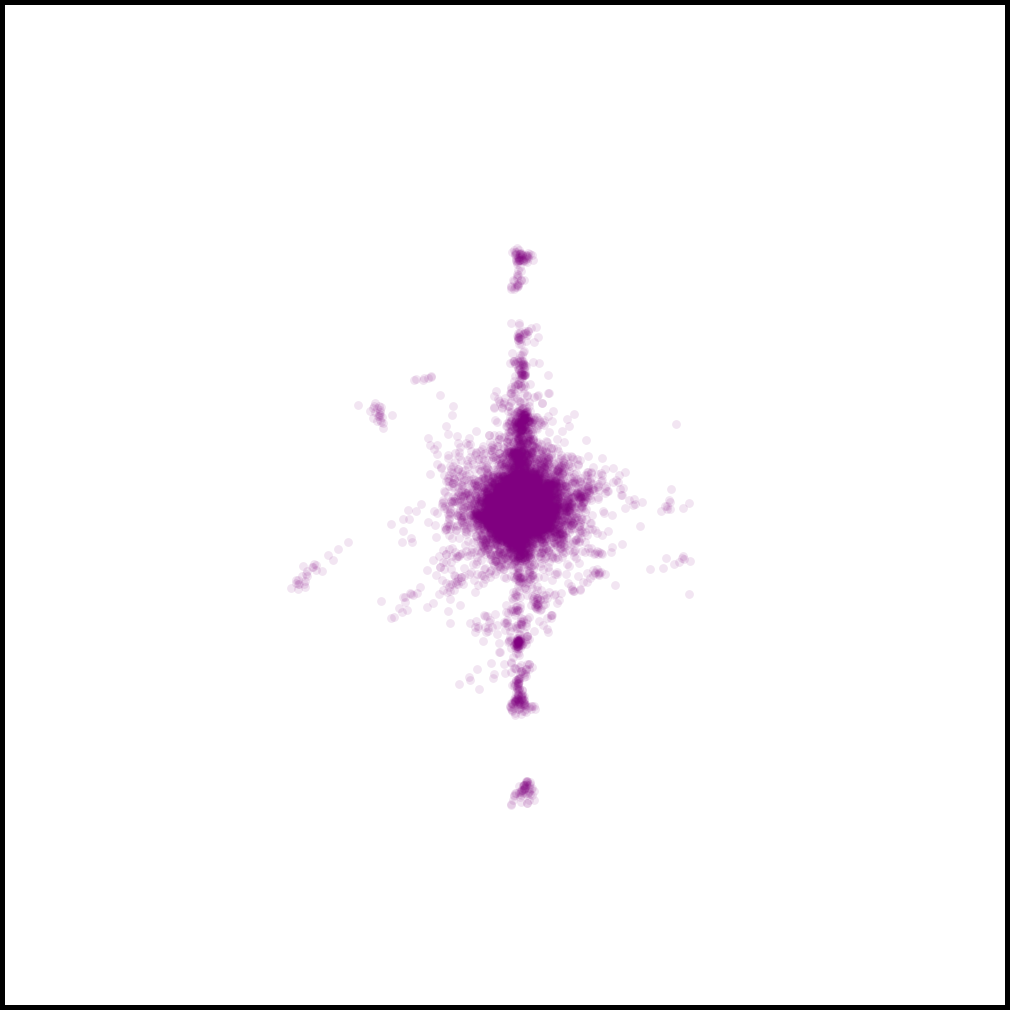} 
  &\includegraphics[width=0.18\linewidth]{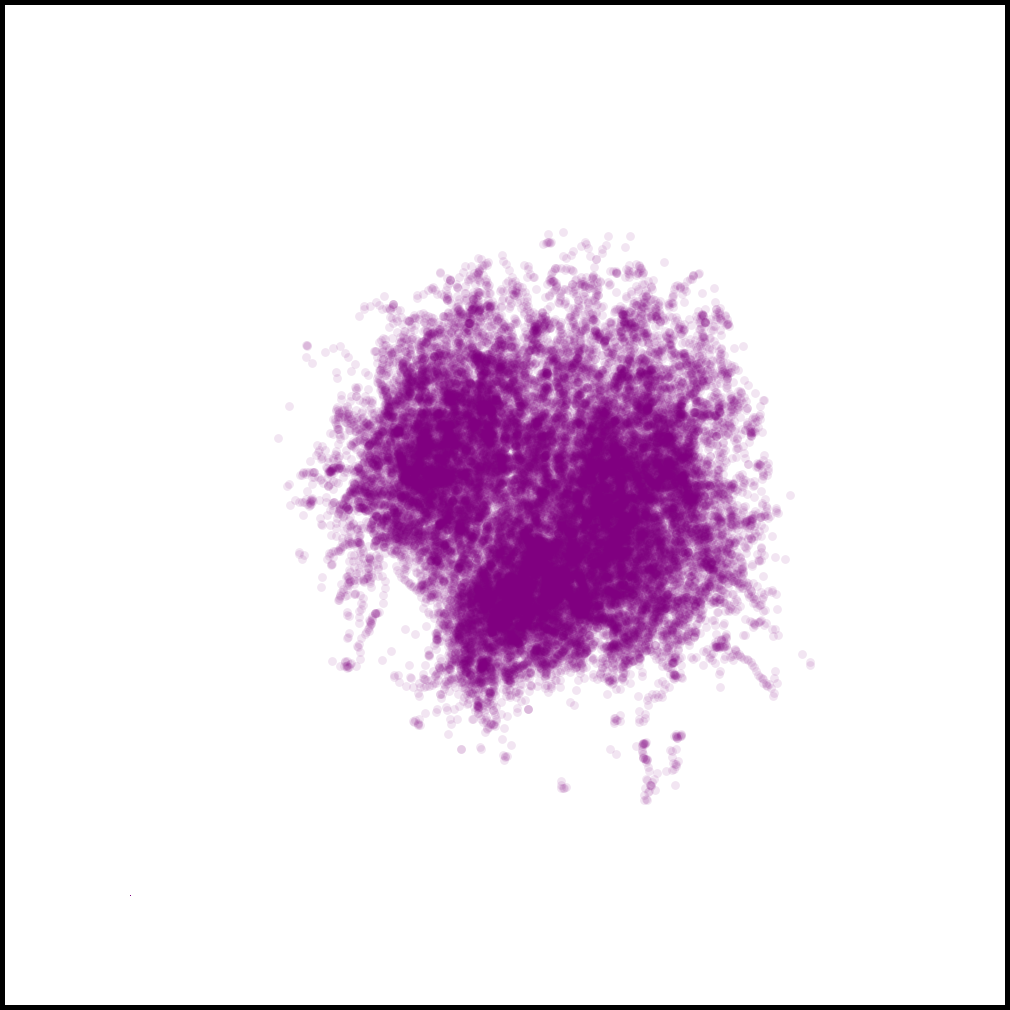} 
  &\includegraphics[width=0.18\linewidth]{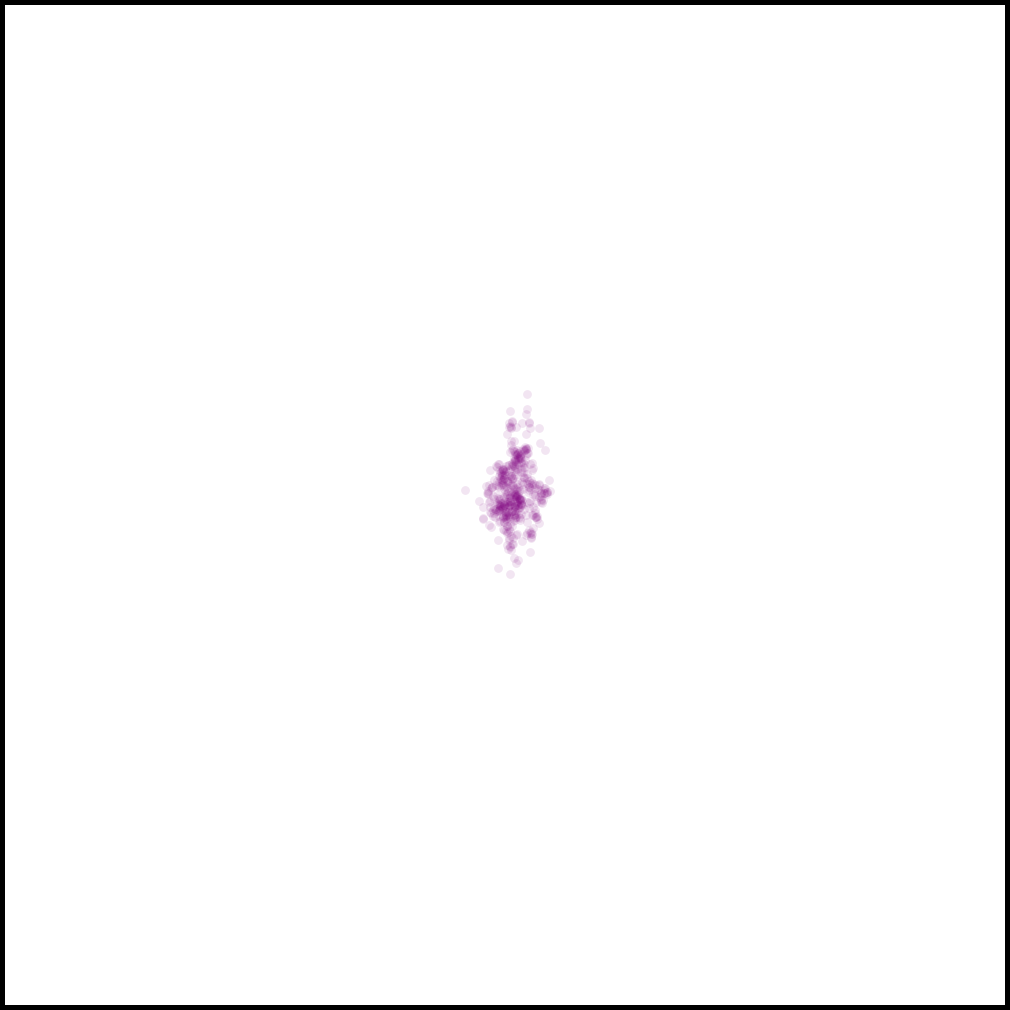} 
  \\
  \end{tabular}
  }
    \caption{Segmentation mask center distributions illustrate center cropping within common medical image segmentation benchmark- and challenge datasets. 
    }
    \label{fig:segmented_object_centre_dist_medical_dataset}
\end{figure}

\textbf{We have shown that shortcut learning is indeed possible for medical image segmentation}, expanding the current discourse which focuses narrowly on classification. We demonstrate two different shortcut mechanisms. The ultrasound case mirrors known shortcut mechanisms from image classification, where models use text and calipers to guide segmentation decisions. Inpainting the calipers can mitigate this. Conversely, the interplay of zero padding and center cropped training sets reveals a different type of shortcut, where modeling choices contribute to changing the feature representations of certain pixels. In this context, we note just how prominent center cropping is in popular benchmarks for medical image segmentation (Fig.~\ref{fig:segmented_object_centre_dist_medical_dataset}): it seems likely that shortcut learning may be a factor in affecting scores in, for instance, segmentation challenges.

\noindent\textbf{Shortcut learning versus overfitting.} Overfitting occurs when excessively complex models capture not just meaningful patterns but also random noise in the training data. This makes the model highly specific to its training set and less capable of generalizing to new data. In contrast, shortcuts lead models to learn superficial and misleading patterns that don't truly represent the underlying problem. The result is still models that perform well on the training data while failing to generalize at test time, but the mechanism is wholly different. 

\noindent\textbf{Domain adaptation does not mitigate shortcut learning.} Domain adaptation implicitly assumes that the patterns learned by the algorithm on the original domain are transferrable to the target domain. In shortcut learning, however, the algorithm has not necessarily learned the underlying patterns at all, because the shortcuts have provided a simpler strategy for optimizing the training loss. As a result, it might not be possible to apply domain adaptation to transfer a model that has learned shortcuts, to data without shortcuts.

\noindent\textbf{Center-cropped shortcuts are beyond CNNs.} Recently popular vision transformers(ViTs)~\cite{dosovitskiy2020image} apply no padding and process image patches, which could mitigate shortcuts as the spatial location of the pixel is no longer included. However, the positional encoding in ViTs might still encode the center bias and lead to a similar shortcut, leaving the shortcut in ViTs an interesting open problem.
  
\noindent\textbf{Beware! Segmentation models are vulnerable to shortcuts.} We aim to highlight that shortcut learning risks extend beyond classification, affecting pixel-level tasks like segmentation, detection, super-resolution, denoising, and artifact removal. Addressing these shortcuts is crucial for developing precise, robust, and reliable machine-learning models for clinical use. We show how simple augmentation and inpainting can mitigate our cases of shortcut learning, while less than $15\%$ of skin lesion segmentation papers since 2014 used random cropping augmentation~\cite{mirikharaji2023survey}, and removing calipers in fetal ultrasound segmentation has not been a standard preprocessing either~\cite{bano2021autofb,zhou2023improving}. Future research should address more generic tools to detect and mitigate shortcuts in image segmentation.

\begin{credits}
\subsubsection{\ackname} This work was supported by the DIREC project EXPLAIN-ME (9142-00001B), the Novo Nordisk Foundation through the Center for Basic Machine Learning Research in Life Science (NNF20OC0062606), and the Pioneer Centre for AI, DNRF grant nr P1. The AI Signature co-funded this work.

\subsubsection{\discintname}
The authors have no competing interests to declare.
\end{credits}

\bibliographystyle{splncs04}
\bibliography{Paper-0423}
\end{document}


\title{Appendix of ``Shortcut Learning in Medical Image Segmentation''}

\titlerunning{Shortcut Learning in Medical Image Segmentation}
\authorrunning{Lin and Weng et al.}
\author{Manxi 
Lin\inst{1\dag}\textsuperscript{\orcidlink{0000-0003-3399-8682}},
Nina Weng\inst{1\dag}\textsuperscript{\orcidlink{0009-0006-4635-0438}}, 
Kamil Mikolaj\inst{1}\textsuperscript{\orcidlink{0000-0002-1631-9329}},
Zahra Bashir\inst{2,3}\textsuperscript{\orcidlink{0000-0002-2497-282X}},
\\ Morten B. S. Svendsen\inst{1,3}\textsuperscript{\orcidlink{0000-0002-4492-3750}},
Martin G. Tolsgaard\inst{3,4,5}\textsuperscript{\orcidlink{0000-0001-9197-5564}},
\\
Anders N. Christensen\inst{1}\textsuperscript{\orcidlink{0000-0002-3668-3128}},
Aasa Feragen\inst{1}\textsuperscript{(\Letter)}\textsuperscript{\orcidlink{0000-0002-9945-981X}}
}

\institute{
$^1$~Technical University of Denmark, Kongens Lyngby, Denmark\\
\email{\{manli, ninwe, afhar\}@dtu.dk}\\
$^2$~Slagelse Hospital, Slagelse, Denmark\\
$^3$~CAMES, Copenhagen, Denmark\\
$^4$ Copenhagen University Hospital Rigshospitalet, Copenhagen, Denmark\\
$^5$ Department of Clinical Medicine, University of Copenhagen, Copenhagen, Denmark
}

\maketitle              

\appendix
\setcounter{page}{1}

\begin{figure}[h]
   \centering
   \includegraphics[width=1.\linewidth]{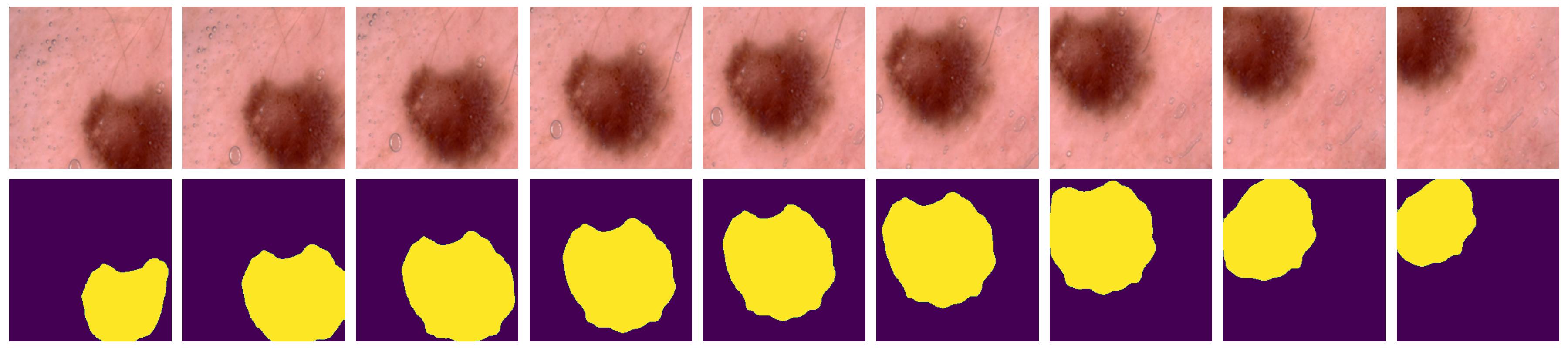}
   \caption{Illustration of how zero padding breaks translation equivariance: When the lesion moves, parts that stay inside the image have unchanged segmentation, whereas parts that are closer to the border are systematically undersegmented.}
   \label{fig:moving_path_sample}
\end{figure}

\begin{figure}[t]
    \centering
  \includegraphics[width=0.9\linewidth]{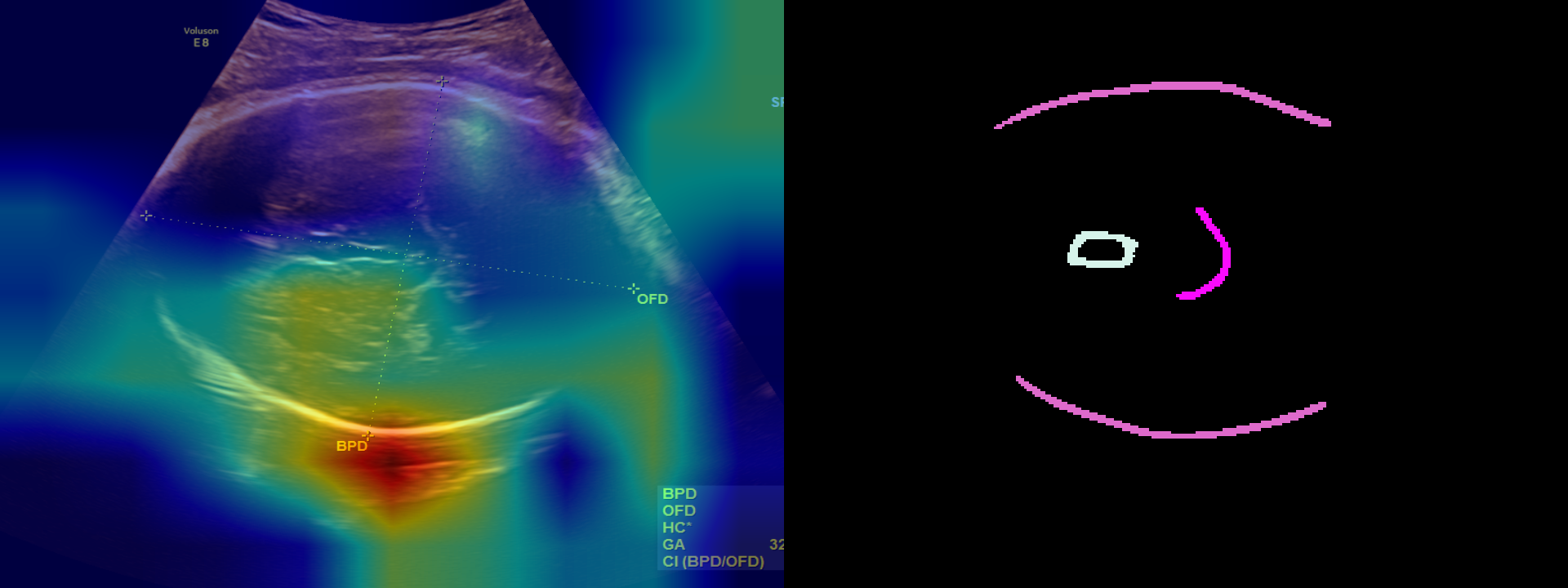} \\
  \includegraphics[width=0.9\linewidth]{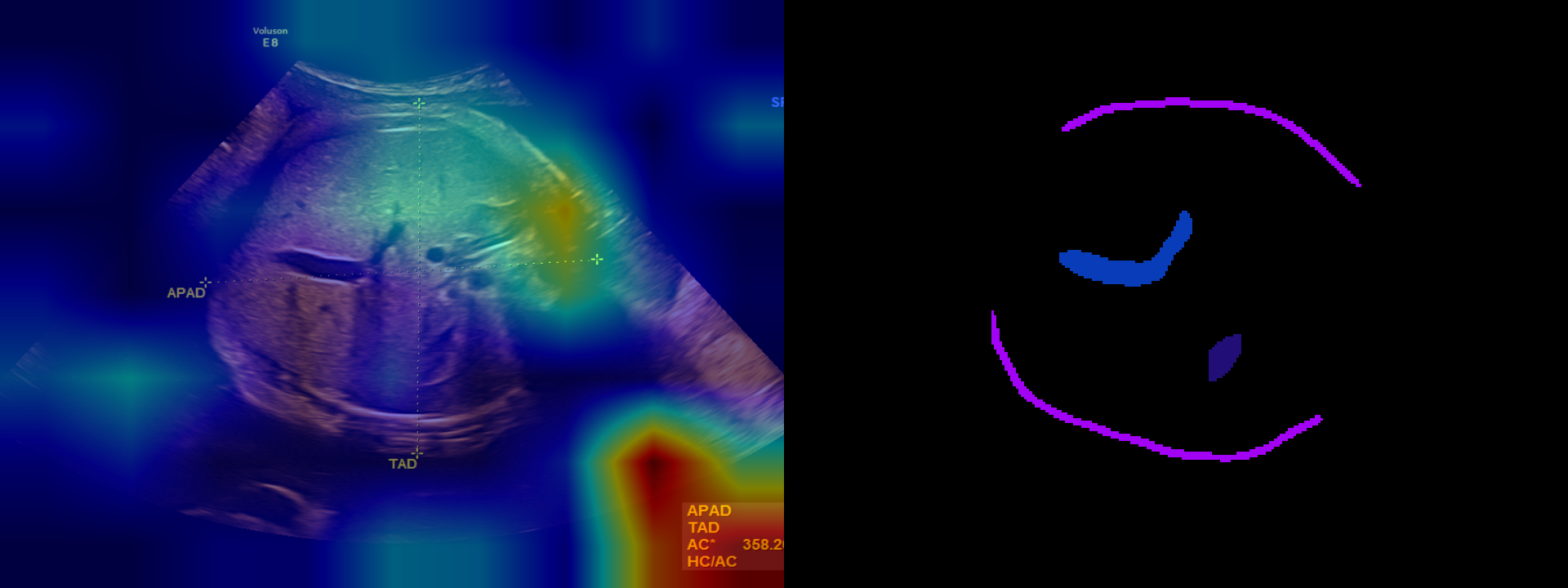} \\
  \includegraphics[width=0.9\linewidth]{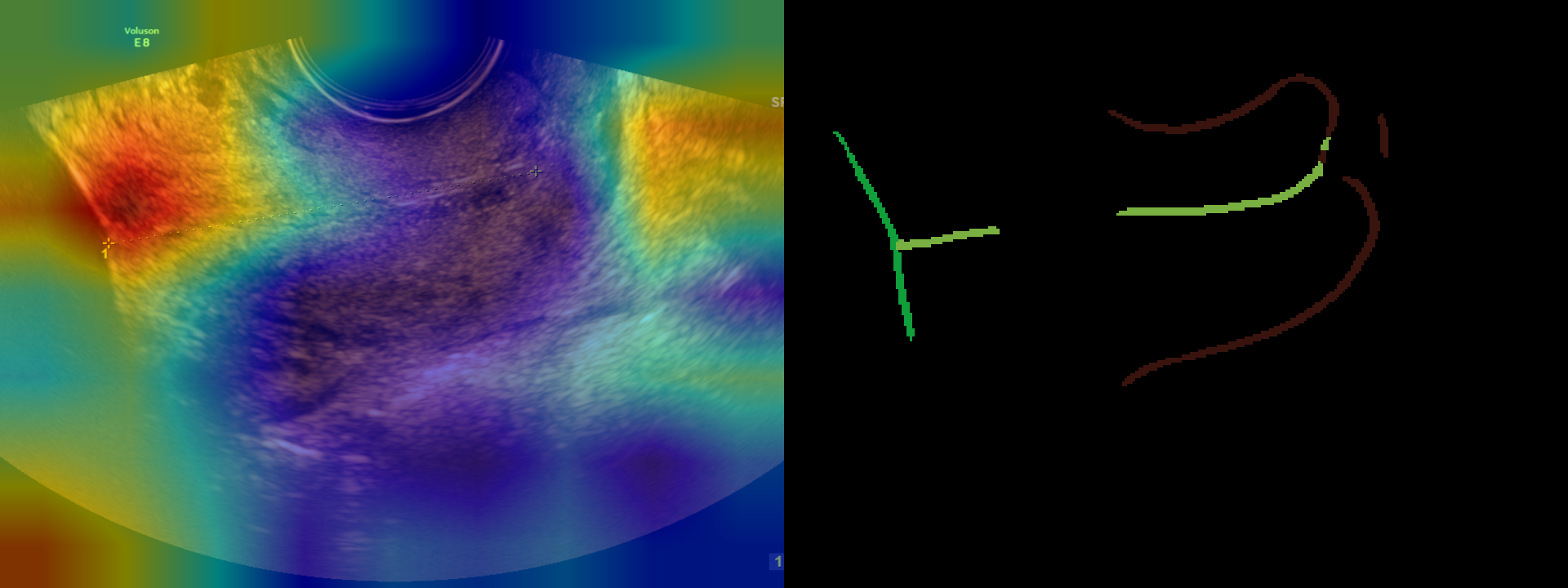} \\
  \includegraphics[width=0.9\linewidth]{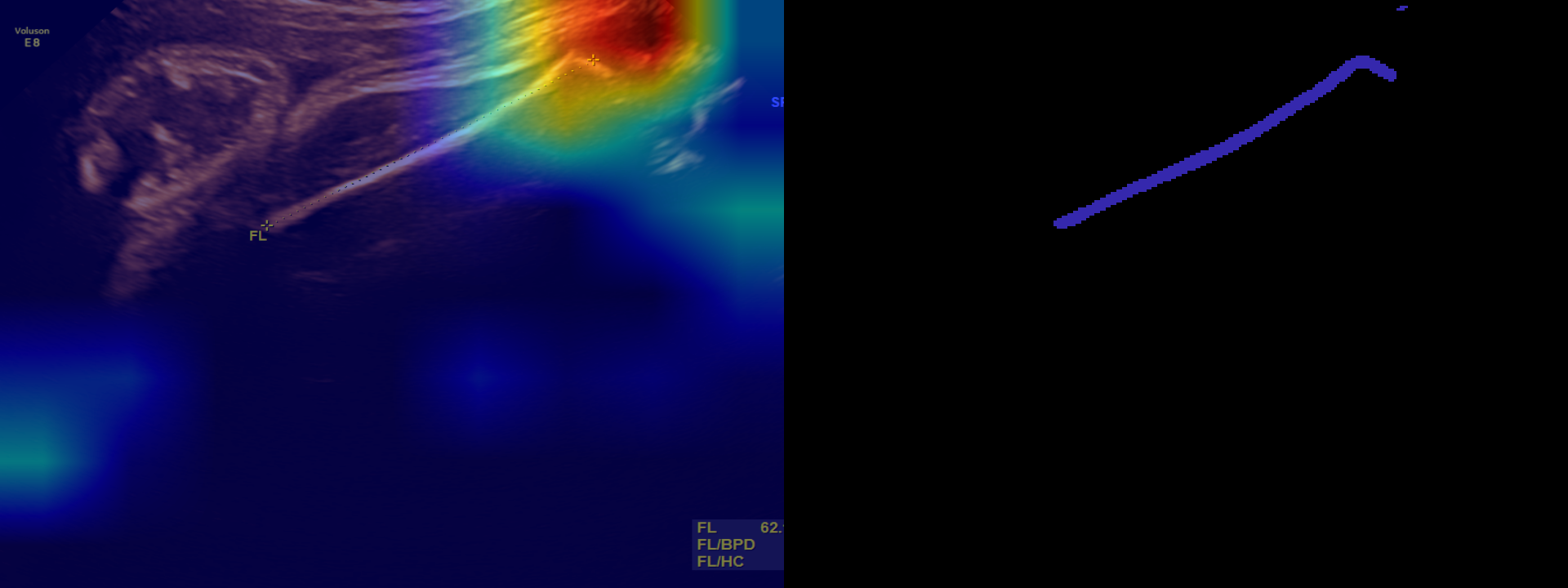} \\
    \caption{Illustration of the saliency maps of the segmentation model on the ultrasound segmentation task. The saliency maps were obtained with Grad-CAM on the last layer of the U-Net encoder. The saliency maps show the model's attention on the proposed shortcuts, i.e., the yellow calipers and texts in the image.
    }
\end{figure}